\documentclass[preprint,12pt]{elsarticle}

\usepackage{amssymb}
\usepackage{color}
\usepackage{microtype}
\usepackage{amsmath}
\usepackage{graphicx}
\usepackage{xcolor}
\usepackage{hyperref}
\usepackage{algorithm}
\usepackage{listings}
\usepackage{tabularx}
\usepackage{soul}
\usepackage{cleveref}
\usepackage{url}
\usepackage{comment}
\usepackage{fancyvrb}
\usepackage{svg}
\usepackage{verbatimbox}
\usepackage[T1]{fontenc}
\usepackage{subcaption}
\usepackage{tcolorbox}
\usepackage[T1]{fontenc}
\usepackage{lmodern} 
\usepackage{algorithmic}

\usepackage{fixltx2e}

\makeatletter
\renewcommand{\ALG@name}{Process Flow} 
\makeatother

\usepackage{tcolorbox}

\newcommand{\numc}[1]{\tikz[baseline=(char.base)]{\node[shape=circle,draw,inner sep=1.1pt] (char) {{\small {#1}}};}}

\usepackage[textsize=scriptsize, 
    ]{todonotes}

\lstdefinestyle{cppstyle}{
    language=C++,
    basicstyle=\ttfamily\footnotesize,
    numbers=left,
    numberstyle=\tiny\color{gray},
    stepnumber=1,
    numbersep=10pt,
    backgroundcolor=\color{white},
    showspaces=false,
    showstringspaces=false,
    showtabs=false,
    frame=none,
    tabsize=2,
    captionpos=b,
    breaklines=true,
    breakatwhitespace=false,
    escapeinside={\%*}{*)},
    xleftmargin=20pt,
    columns=flexible,
    keywordstyle=\bfseries\color{green!40!black},
    commentstyle=\itshape\color{purple!40!black}
}

\journal{Science of Computer Programming}

\begin{document}

\begin{frontmatter}



\title{ESBMC v7.6: Enhanced Model Checking of C++ Programs with Clang AST}


\author[label1]{Xianzhiyu Li\corref{cor1}}
\ead{xianzhiyu.li@postgrad.manchester.ac.uk}

\author[label1]{Kunjian Song}
\ead{kunjian.song@postgrad.manchester.ac.uk}

\author[label2]{Mikhail R. Gadelha}
\ead{mikhail@igalia.com}

\author[label1]{Franz Brau\ss e}
\ead{franz.brausse@manchester.ac.uk}

\author[label1,label3]{Rafael S. Menezes}
\ead{rafael.menezes@postgrad.manchester.ac.uk}

\author[label1]{Konstantin Korovin}
\ead{konstantin.korovin@manchester.ac.uk}

\author[label1,label3]{Lucas C. Cordeiro}
\ead{lucas.cordeiro@manchester.ac.uk}

\cortext[cor1]{Corresponding author}
\affiliation[label1]{organization={The University of Manchester},
            addressline={Oxford Rd}, 
            city={Manchester},
            postcode={M13 9PL}, 
            state={England},
            country={UK}}

\affiliation[label2]{organization={Igalia},
            addressline={Bugallal Marchesi, 22, 1º}, 
            city={A Coruña},
            postcode={15008}, 
            state={Galicia},
            country={Spain}}

\affiliation[label3]{organization={Federal University of Amazonas},
            addressline={Av. General Rodrigo Oct\'{a}vio Jord\~{a}o Ramos, 6200}, 
            city={Manaus},
            postcode={69080-005}, 
            state={Amazonas},
            country={Brazil}}

\begin{abstract}
This paper presents Efficient SMT-Based Context-Bounded Model Checker (ESBMC) v7.6, an extended version based on previous work on ESBMC v7.3 by K. Song et al.~\cite{song2023esbmc}. The v7.3 introduced a new Clang-based C++ front-end to address the challenges posed by modern C++ programs. Although the new front-end has demonstrated significant potential in previous studies, it remains in the developmental stage and lacks several essential features. ESBMC v7.6 further enhanced this foundation by adding and extending features based on the Clang AST, such as  \numc{1} exception handling, \numc{2} extended memory management and memory safety verification, including dangling pointers, duplicate deallocation, memory leaks and rvalue references and \numc{3} new operational models for STL updating the outdated C++ operational models. Our extensive experiments demonstrate that ESBMC v7.6 can handle a significantly broader range of C++ features introduced in recent versions of the C++ standard.
\end{abstract}




\end{frontmatter}


\section{Introduction}
\label{intro}

C++ is one of the most popular programming languages used to build high-performance and real-time systems, such as operating systems, banking systems, communication systems, and embedded systems~\cite{deitel2014c++}. However, memory safety issues remain a major source of security vulnerabilities in C++ programs~\cite{miller2019trends}. Fan et al.~\cite{fan2020ac} created a dataset of C/C++ vulnerabilities by mining the Common Vulnerabilities and Exposures (CVE) database~\cite{cvemitreurl} and the associated open-source projects on GitHub, then curated the issues based on Common Weakness Enumeration (CWE)~\cite{cwemitreurl}. According to their findings, two out of the top three vulnerabilities are caused by memory safety issues: Improper Restriction of Operations within the Bounds of a Memory Buffer (CWE-119) and Out-of-bounds Read (CWE-125)~\cite{fan2020ac}.

The limitation of software testing resides in the user inputs~\cite{quadri2010software}. Only a limited number of execution paths may be tested since test cases involve human inputs in the form of concrete values~\cite{ammann2016introduction}. In contrast to testing, formal verification techniques can be used more systematically to formally reason about a program, although they suffer from the state-space explosion problem~\cite{MonteiroGCF17}. There is an increasing adoption of formal verification techniques for C programs in the industry, e.g., Amazon has been using model-checking techniques to prove the correctness of their C-based systems in Amazon Web Services (AWS); this has positively impacted their code quality, as evidenced by the increased rate of bugs found and fixed~\cite{chong2020code}. 

Formal verification of C++ programs is more challenging than C programs due to the sophisticated features, such as the STL (Standard Template Libraries) containers, templates, exception handling, and object-oriented programming (OOP) paradigm~\cite{deitel2014c++}. Several tools have been developed for verification of C++ programs, most prominent are: CBMC~\cite{clarke2004tool}, DIVINE~\cite{baranova2017model}, and ESBMC~\cite{song2023esbmc}. 
But the existing state-of-the-art verification tools have only limited C++ feature support~\cite{monteiro2022model}. 
CBMC~\cite{clarke2004tool} is a bounded model checker that converts source code into an intermediate representation for verification, with limited support for polymorphism and exception handling~\cite{cordeiro2011smt}. DIVINE~\cite{baranova2017model}, an explicit-state model checker, uses LLVM bitcode~\cite{clangllvm} as an intermediate representation for verifying C++ programs, with limited ability to handle standard containers and inheritance~\cite{monteiro2022model}.
For ESBMC, Ramalho et al.~\cite{ramalho2013smt} and Monteiro et al.~\cite{monteiro2022model} initiated the support for C++ program verification. Since then, ESBMC has undergone heavy development to support recent versions of the C++ standard~\cite{cpp20std}.
This research builds upon the Clang-based C++ front-end introduced in ESBMC v7.3, which replaced the legacy parser with a more robust tool. In this work, we refine and extend these contributions:

\begin{itemize}
\item \textbf{Complete Redesign}: ESBMC's C++ front-end has undergone a complete restructuring and now relies on Clang~\cite{clangllvm}. By leveraging Clang's parsing and semantic analysis capabilities~\cite{lopes2014getting,pandey2015llvm}, we check the input program's Abstract Syntax Tree (AST) using a production-quality compiler. This eliminates static analysis logic and ensures enhanced accuracy and efficiency.
\item \textbf{Object Models Details}: We provide comprehensive insights into the object models used to achieve seamless conversion of C++ polymorphism code to ESBMC's Intermediate Representation (IR). 
\item \textbf{Simplified Type Checking for Templates:} The new Clang-based front-end greatly simplifies type checking for templates, streamlining ESBMC's ability to adapt to C++ advancements. 
\end{itemize}

Between ESBMC v7.3 and this work, we not only refined existing features but also introduced several new verification capabilities:
\begin{itemize}
\item \textbf{Extended C++ Memory Management:} We have extended the implementation of the dynamic memory operators \texttt{new} and \texttt{delete} in our new front-end, which enhances ESBMC's ability to verify memory safety issues.
\item \textbf{Modeled Rvalue References:} Our new Clang-based C++ front-end models the key C++11 feature of \textit{rvalue} references~\cite{josuttis2012c++}, supports the \textit{move} function and \textit{move} semantics.
\item \textbf{C++ Exception Handling:} We have implemented exception handling based on Clang AST and extended the exception specification. Additionally, we adapted the symbolic engine to match thrown exceptions. This enhancement enables ESBMC to support exceptions in C++11 and later versions. 
\item \textbf{Updated C++ Operational Models (OMs):} We enabled OMs from ESBMC v2.1 and maintained the outdated OMs to adapt them for the new Clang-C++ front-end.
\end{itemize}

By introducing these features, our work significantly enhances ESBMC's C++ verification capabilities, paving the way for more robust and efficient verification of C++ programs and their variants. This work primarily focuses on the improvements made to ESBMC across its different versions, with contributions being relative to its previous iterations. Comparisons with other verification tools will be explored in future work.

This paper is organized as follows: we begin with a brief introduction to SMT-based BMC techniques and the limitations of previous versions of ESBMC. In Section~\ref{clang-based approach}, we present the implementation of core C++ language features based on the Clang AST, along with detailed process flows. Section~\ref{experimental-evaluation} provides the experimental results of the benchmarks used and analyzes potential threats to validity. Finally, in Section~\ref{conclusion-and-future-work}, we conclude and outline future work.

\section{Background}
\label{background}

ESBMC's verification for C++03 programs reached its maturity in version v2.1, presented by Monteiro et al.~\cite{monteiro2022model}. ESBMC v2.1 provides a first-order logic-based framework that formalizes a wide range of C++ core languages, verifying the input C++ programs by encoding them into SMT formulas. Since C++ Standard Template Libraries (STL) contain optimized assembly code not verifiable using ESBMC, ESBMC v2.1 tackled this problem using a collection of C++ Operational Models (OMs) to replace the STL included in the input program. The OMs are abstract representations mimicking the structure of the STL, adding pre- and post-conditions to all STL APIs~\cite{dos2015simple}. Combining these approaches, ESBMC v2.1 outperformed other state-of-the-art tools evaluated over a large set of benchmarks, comprising $1513$ test cases~\cite{monteiro2022model}. Nonetheless, ESBMC v2.1 employs a Flex and Bison-based front-end from CBMC~\cite{clarke2004tool}, which leads to hard-to-maintain code and can hardly evolve to support modern features introduced in C++11 and later versions. Between v2.1 and v7.3, the focus was on developing a new C++ front-end. C++ verification using Clang AST was introduced in v7.3, which significantly improved the Clang-based front-end and made it the first version capable of handling most modern C++ features. As a result, v2.1 serves as the meaningful point of comparison, as it represents the state of ESBMC before the new C++ front-end was introduced.
\subsection{SMT-based BMC technique}
\label{smt}

The core functionality of ESBMC is based on SMT solvers to process a decidable fragment of first-order logical formulas derived from intermediate representations (IR), thereby enabling efficient model checking. In BMC, the analyzed program is modeled as a state transition system, derived from the control-flow graph (CFG)~\cite{muchnick1997advanced}. The CFG is created during the translation from program code to single static assignment (SSA) form. Nodes in the CFG represent assignments or conditional statements, while edges represent possible changes in the program's control flow. Consider a transition system \( M \), a property \( \phi \), and an integer parameter \( k \). Bounded Model Checking (BMC) unrolls the system \( k \) times, converting it into a verification condition \( \psi_k \). The condition \( \psi_k \) is satisfiable if and only if a counterexample of length \( k \) or less exists for the property \( \phi \). This model-checking problem can be formalized by constructing the following logical formula:

\begin{equation}
\psi_k = I(s_0) \land \bigwedge_{i=0}^{k-1} T(s_i, s_{i+1}) \land \bigvee_{i=0}^{k} \neg \phi(s_i)
\label{eq:bmc_formula}
\end{equation}

In this formulation, \( I \) denotes the set of initial states of \( M \), \( T(s_i, s_{i+1}) \) describes the transition relation between states \( s_i \) and \( s_{i+1} \), and \( \phi(s_i) \) is the safety property evaluated at state \( s_i \). The formula \( I(s_0) \land \bigwedge_{i=0}^{k-1} T(s_i, s_{i+1}) \) represents the executions of \( M \) over \( k \) steps, while \( \bigvee_{i=0}^{k} \neg \phi(s_i) \) indicates that \( \phi \) is violated in some state \( s_i \) for \( 0 \leq i \leq k \).  
In cases where the formula~\eqref{eq:bmc_formula} is satisfiable, an SMT solver can produce a satisfying assignment. This assignment enables us to determine the values of the program variables, which can then be used to construct a counterexample. Such a counterexample for the property \( \phi \) is a sequence of states \( s_0, s_1, \ldots, s_k \) where \( s_0 \in S_0 \), with \( S_0 \) representing the set of initial states, and \( T(s_i, s_{i+1}) \) holds for \( 0 \leq i < k \).

On the other hand, if the formula~\eqref{eq:bmc_formula} is unsatisfiable, no error state is reachable in \( k \) steps or fewer. However, this does not guarantee the completeness of BMC techniques, as counterexamples with lengths exceeding \( k \) may still exist. To ensure completeness, it is necessary to establish an upper bound on the depth of the state space. This involves confirming that all significant system behaviors have been explored, ensuring that further search only exhibits states that have already been verified, which can be achieved using k-induction to prove that the system remains correct beyond the explored states~\cite{kroening2011linear, gadelha2019esbmc}.

\subsection{Old C++ front-end in v2.1}
\label{limitations-of-old-frontend}
The version of ESBMC in Monteiro et al.~\cite{monteiro2022model} used an outdated CPROVER-based front-end~\cite{clarke2004tool} with the following limitations.

\begin{enumerate}
  \item For the type-checking phase, ESBMC could not provide meaningful warnings or error messages.
  \item It was inefficient at generating a body for default implicit non-trivial methods in a class, such as C++ copy constructors or copy assignment operators. 
  \item The parser of the old front-end needed to be manually updated to cover the essential C++ semantic rules~\cite{esbmccpptypecheck}, which leads to hard-to-maintain code to keep up with the C++ evolution. 
  \item The old front-end contained excessive data structures and procedures auxiliary to scope resolution and function type-checking.
  \item The type-checker~\cite{esbmccpptypecheck} of the old front-end only worked with a CPROVER-based parse tree and supports features up to the C++03 standard~\cite{cpp03std}. We found adapting it to the new C++ language and library features difficult. 
  \item The old front-end used a speculative approach to guess the arguments for a template specialization and a map to associate the template parameters to their instantiated values, which leads to hard-to-maintain and hard-to-debug code in the case of recursive templates. Additionally, owing to its limited static analysis, the old front-end could not provide any early warning when there is a circular dependency between the templates.
\end{enumerate}

Following the introduction of ESBMC v7.3 by K. Song et al.~\cite{song2023esbmc}, these limitations have been addressed. Clang can provide detailed error messages and warnings during the type-checking phase and automatically generate implicit methods for classes. Notably, the Clang-based approach solves the problem of needing continuous maintenance to adapt to new C++ feature changes. However, ESBMC v7.3 lacked full support for \textit{rvlaue} references, exception handling, and dynamic memory verification. These limitations affected its ability to handle modern C++ programs. In this work, we extend the Clang-based C++ front-end by addressing these gaps, proposing approaches to ensure compliance with recent C++ standards~\cite{cpp20std}.

\section{Model Checking C++ Programs using Clang AST}
\label{clang-based approach}

\begin{figure*}[htbp]
  \centering
  \includegraphics[width=0.9\textwidth]{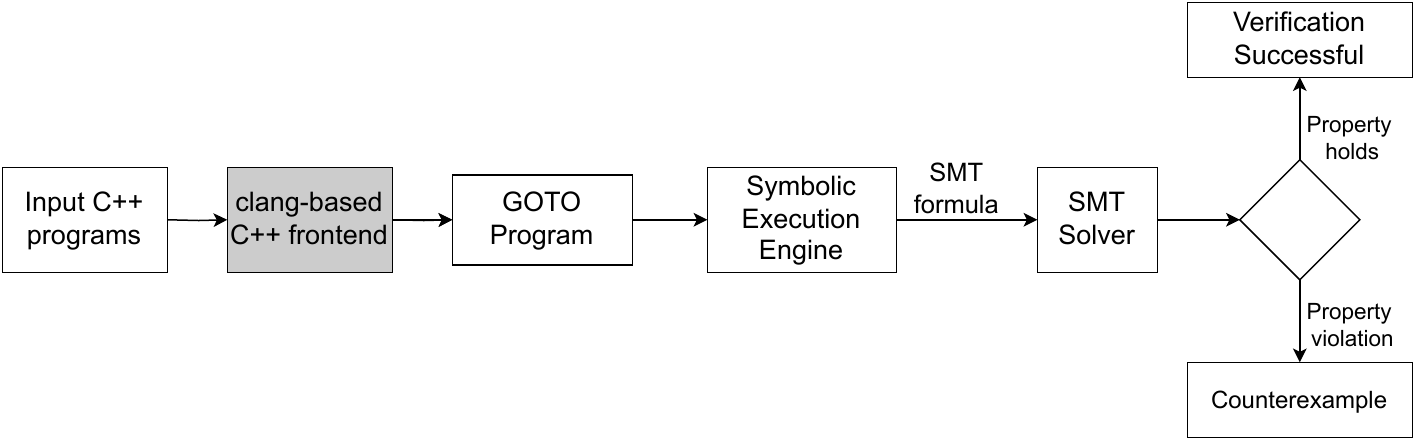}
  \caption{ESBMC architecture for C++ verification. The grey block represents the extended Clang-based C++ front-end developed in ESBMC v7.6.}
  \label{fig:esbmcpp-architecture}
\end{figure*}

Figure~\ref{fig:esbmcpp-architecture} illustrates ESBMC's verification pipeline for C++ programs. The new Clang-C++ front-end type-checks and converts the input C++ program (along with the corresponding OMs) into the GOTO program representation~\cite{cordeiro2011smt,CordeiroF11}. Then, the GOTO program will be symbolically executed to generate the SSA form of the program, thus generating a set of logical formulas consisting of the constraints and properties. An SMT solver is used to check the satisfiability of the formulas, giving a verdict \textit{VERFICATION SUCCESSFUL} if no property violation is found up to the bound $k$ or a counterexample in case of property violation (cf. Section~\ref{smt}).

\subsection{Polymorphism}
\label{polymorphism}

The traditional approach for achieving polymorphism makes use of virtual function tables (also known as \textit{vtables}) and virtual pointers (known as \textit{vptrs}). While the Clang AST does not include information about virtual tables or virtual pointers of a class, it provides users with enough information to enable them to create their own \textit{vtables} and \textit{vptrs}. In the new Clang-based C++ front-end, we reimplemented the \textit{vtable} and \textit{vptr} construction mechanism following a similar approach as ESBMC v2.1, but with significant simplifications based on the information provided in the Clang AST. Figure~\ref{fig:cpp_poly} illustrates an example of C++ polymorphism.

\begin{figure}[htbp]
  \centering
    \begin{subfigure}[c]{.6\textwidth}
    \begin{lstlisting}[style=cppstyle]
class Bird {
  public:
  virtual int doit(void) { return 21; }
};

class Penguin: public Bird {
  public:
  int doit(void) override { return 42; }
};

int main() {
  Bird *p = new Penguin();
  assert(p->doit() == 42);
  delete p;
  return 0;
}
    \end{lstlisting}
    \end{subfigure}
\caption{Example of C++ classes with virtual functions.}
\label{fig:cpp_poly}
\end{figure}

\begin{figure}[htbp]
  \centering
  \includegraphics[width=1\textwidth]{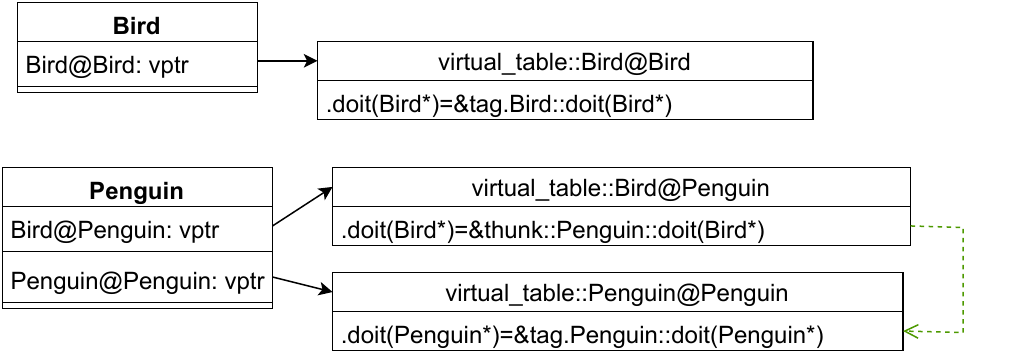}
  \caption{Object models for Bird and Penguin classes}
  \label{fig:cpp_poly_objmodel}
\end{figure}

\begin{figure}[htbp]
  \centering
    \begin{subfigure}[c]{.4\textwidth}
    \vspace{0.5cm}
    \begin{lstlisting}[style=cppstyle]
int return_value;
return_value =
*p->Bird@Penguin
    ->doit(p)
assert(return_value == 42)
    \end{lstlisting}
    \caption{GOTO program of the dynamic dispatch in Line 12 of Figure~\ref{fig:cpp_poly}.}
    \label{fig:cpp_poly_goto_sub1}
    \end{subfigure}
    \hspace{1.cm}
    \begin{subfigure}[c]{.45\textwidth}
    \begin{lstlisting}[style=cppstyle]
thunk::Penguin::doit(Bird*):
  int return_value;
  return_value =
    Penguin::doit(
      (Penguin*)this)
  RETURN: return_value
  END_FUNCTION

Penguin::doit(Penguin*):
  RETURN: 42
  END_FUNCTION
    \end{lstlisting}
    \caption{Thunk redirecting the call to the overriding function.}
    \label{fig:cpp_poly_goto_sub2}
    \end{subfigure}
\caption{GOTO conversions of the overriding methods and dynamic dispatch.}
\label{fig:cpp_poly_goto}
\end{figure}

Figure~\ref{fig:cpp_poly_objmodel} illustrates the object models for the example classes \lstinline[style=cppstyle]!Bird! and \lstinline[style=cppstyle]!Penguin!. The new front-end adds one or more \textit{vptrs} to each class. The \textit{vptrs} will be initialized in the class constructors, which set each \textit{vptr} pointing to the desired \textit{vtable}. The child class contains an additional pointer pointing to a \textit{vtable} with a thunk to the overriding function. The thunk redirects the call to the corresponding overriding function. In case of multiple inheritances, the child class would have multiple \textit{vtprs} ``inherited'' from multiple base classes. The new front-end can also manage virtual inheritance, such as the diamond problem, which avoids duplicating \textit{vptrs}, referring to the same virtual table in an inheritance hierarchy. Lines 2-4 in Figure~\ref{fig:cpp_poly_goto_sub1} illustrate how dynamic dispatch is achieved using the \textit{vptr} calling the thunk, which in turn calls the desired overriding function in Figure~\ref{fig:cpp_poly_goto_sub2}, lines 9-11. Note that the \textit{override} specifier is a C++11 extension that the old front-end could not support. As shown in Process Flow~\ref{pf:poly}, class definitions are parsed from Clang AST to establish derived and base class relationships. An object model is created with \textit{vptrs} to the \textit{vtable}. Method calls are redirected through \textit{vtpr} to overriding functions, or the base class function is invoked if no override exists.

\begin{algorithm}[htbp]
\footnotesize
\algsetup{linenosize=\footnotesize}
\caption{\footnotesize Polymorphism}
\label{pf:poly}
\begin{algorithmic}[1]
\STATE Parse the class definition from Clang AST to get the derived class and its base class

\STATE Create an \textit{object model} for classes, storing each \textit{vtpr} 
that point to the \textit{vtable}.

\IF{Use \textit{vtpr} call thunk virtual functions}
    \STATE Redirect the call to the corresponding overriding function
\ELSE
    \STATE Call the base class function
\ENDIF

\end{algorithmic}
\end{algorithm}
\subsection{Templates}
\label{template} 

Templates are a key feature in C++, allowing types and certain values to be passed as parameters to types, values and functions. Templates allow STL containers and generic algorithms to work with different C++ data types~\cite{prata2012c++,stroustrup2013c++}. The old front-end in ESBMC v2.1 implements template specialization based on Siek et al.~\cite{siek2006semantic,monteiro2022model}. However, it produces a ``CONVERSION ERROR'' for the test case illustrated in Figure~\ref{fig:template_example_code}. This benchmark is based on the \textit{Friend18} example from the GCC test suite~\cite{gcctestsuite}, which was added for Bug $10158$ on GCC Bugzilla~\cite{gcc10158}. ESBMC v7.6 successfully verified this benchmark and found the assertion's property violation in Figure~\ref{fig:template_example_code}. The verification result is illustrated in Figure~\ref{fig:tempalte_verfication_verdict}. The example in Figure~\ref{fig:template_example_code} contains a C++20 extension. The \textit{foo} function is defined in \textit{struct X} but gets called using an unqualified name with explicit template arguments in \textit{main}. ESBMC v2.1 failed to verify it due to the ``CONVERSION ERROR symbol `foo' not found''. 
We also tried this example with CBMC 5.88.1~\cite{cbmc5881rel} and cppcheck v2.11.1~\cite{cppcheck2111rel}, both current as of the time of writing. CBMC aborted during type-checking, while cppcheck did not provide any verification verdict. The Process Flow~\ref{pf:template} shows how the Clang-based C++ front-end parses and processes templates.

\begin{figure*}[htbp]
  \centering
    \begin{subfigure}[c]{.45\textwidth}
    \begin{lstlisting}[style=cppstyle]
#include <cassert>
template <int N> struct X
{
  template <int M>
  friend int foo(X const &)
  {
    return N * 10000 + M;
  }
};
X<1234> bring;

int main() {
  assert(
   foo<5678> (bring)
    !=12345678);
}
    \end{lstlisting}
    \caption{Example of C++ class template}
    \label{fig:template_example_code}
    \end{subfigure}
    \hspace{1cm}
    \begin{subfigure}[c]{0.4\textwidth}
    \begin{lstlisting}[style=cppstyle]
Violated property:
  file tmp2.cpp 
   line 13 column 3 
    function main
  assertion 
   foo<5678>(bring)!=12345678
  return_value!=12345678

VERIFICATION FAILED
    \end{lstlisting}
    \caption{Verdict for the template example}
    \label{fig:tempalte_verfication_verdict}
    \end{subfigure}
\caption{ESBMC verified the \textit{Friend18} example from the GCC test suite.~\cite{gcctestsuite}}
\label{fig:cpp_template_example}
\end{figure*}

\begin{algorithm}[htbp]
\footnotesize
\algsetup{linenosize=\footnotesize}
\caption{\footnotesize Templates}
\label{pf:template}
\begin{algorithmic}[1]
\STATE Parses Template Class definition and Friend Function Templates defined in it

\STATE Instantiate the Template Class and Friend Function Template

\STATE Convert Clang AST into an intermediate representation (IR)

\end{algorithmic}
\end{algorithm}

\subsection{C++ New and Delete}
\label{new}

\emph{New} and \emph{Delete} are key operators for dynamic memory management in C++. \emph{New} is used to dynamically allocate memory space and assign a pointer to that space to a variable, \emph{Delete} is used to release the memory allocated by \emph{New}. Note that this can lead to memory safety issues if used incorrectly~\cite{MonteiroGCF17}; for example, if a pointer is not set to \emph{nullptr} after releasing the memory, it is a dangling pointer. When the program continues to use this dangling pointer, it may lead to undefined behavior. ESBMC v7.6 can verify such memory safety issues and the asserted property violation gives accurate information as shown in Figure~\ref{fig:new_result}.

\begin{figure*}[htbp]
  \centering
    \begin{subfigure}[c]{.4\textwidth}
    \begin{lstlisting}[style=cppstyle]
class Foo {
  public:
    Foo() {value = 0;};
    void Inc() {value++;};
  private:
    int value;
};

int main() {
  Foo *foo = new Foo();
  delete foo;
  foo->Inc();
  return 0;
}
    \end{lstlisting}
    \caption{Example of C++ new and delete}
    \label{fig:new_code}
    \end{subfigure}
    \hspace{1cm}
    \begin{subfigure}[c]{0.45\textwidth}
    \begin{lstlisting}[style=cppstyle]
Violated property:
 file main6.cpp
  line 4 column 17
   function Inc
 dereference failure:
  invalidated dynamic object

VERIFICATION FAILED
    \end{lstlisting}
    \caption{Verdict for the new example}
    \label{fig:new_result}
    \end{subfigure}
\caption{ESBMC verified an example with new and delete}
\label{fig:cpp_new_example}
\end{figure*}

ESBMC v7.6 employs a flat memory model and treats dynamically allocated memory as a contiguous region. It supports dynamic arrays and allows pointer arithmetic, with strict bounds checking to prevent out-of-bounds access while allowing unbounded array sizes. Built-in arrays track variables that point to dynamic memory, including their size and validity; each dynamic object is marked as invalid when deallocated using \emph{delete}. It ensures the validity of dynamic objects when accessed and modified, enabling it to verify a wide range of memory safety issues, such as duplicate deallocation and memory leaks. We also extend the implementation of \emph{delete} to check whether the correct operator is used based on the size. For example, memory allocated for dynamic arrays should be released using the \emph{delete} array operator, which is not supported in ESBMC v2.1~\cite{monteiro2022model}.

The Process Flow~\ref{pf:new} begins by parsing the \emph{new} and \emph{delete} operators and the object from the Clang AST. The object is initialized, and a new dynamic object is created and marked as valid. When the \emph{delete} operator is invoked, the memory allocation is verified for consistency, and the dynamic object is marked as invalid. During dereferencing operations, the validity of the dynamic object is checked to ensure safe access.

\begin{algorithm}[htbp]
\footnotesize
\algsetup{linenosize=\footnotesize}
\caption{\footnotesize C++ \emph{New} and \emph{Delete}}
\label{pf:new}
\begin{algorithmic}[1]
\STATE Parse the \emph{New}, \emph{Delete} operators and object via Clang AST

\STATE Initialize the object (e.g., call the constructor)

\STATE Create a \emph{dynamic object} for it and mark it as \emph{valid}

\IF{Call the \emph{Delete} operator to free the memory allocated}
    \STATE Check if the operator matches (e.g., new[] and delete[])
    \STATE Mark the corresponding \emph{dynamic object} as \textit{invalid}
\ENDIF

\IF{Dereference the \emph{dynamic object}}
    \STATE Check if the \emph{dynamic object} is valid
\ENDIF

\end{algorithmic}
\end{algorithm}

\subsection{Rvalue References}
\label{rvalue}

Rvalue references were introduced in C++11, enhancing the efficiency of object operations and its syntax is to append double \& after the type~\cite{josuttis2012c++}. As the foundation of move semantics, it allows transferring resources from temporary objects to another object, eliminating the need for costly deep copy operations. This greatly enhances the efficiency of object assignment and passing, especially for large objects and dynamically allocated resources.

\subsubsection{Move Semantics}

ESBMC v7.6 models \textit{rvalue} references and \textit{move} semantics based on the information provided by Clang AST; we also supported the \textit{move} function, which is used to convert the parameter into an \textit{rvalue} reference explicitly. As shown in Process Flow~\ref{pf:rvalue}, we model \textit{rvalue} references as pointers, which are dereferenced when assigned an \textit{rvalue} or involved in computations with \textit{rvalues}, including boolean operations. The example provided in Figure~\ref{fig:rvalue_code} involves the usage of the \textit{move} function and the assignment to \textit{rvalue} reference. Figure~\ref{fig:rvalue_goto} illustrates our Clang-based C++ front-end's modeling of \textit{rvalue} reference and the special treatment given to their operations. Assertions are used to ensure the validity of each step of these operations.

\begin{figure*}[htbp]
  \centering
    \begin{subfigure}{.45\textwidth}
    \begin{lstlisting}[style=cppstyle]
#include <cassert>
#include <utility>
int main() {
    int a = 10;
    int &&rref = std::move(a);
    assert(rref == 10);
    rref = 5;
    assert(rref == 5);
    return 0;
}
    \end{lstlisting}
    \caption{Example of Rvalue reference}
    \label{fig:rvalue_code}
    \end{subfigure}
    \hspace{1cm}
    \begin{subfigure}{0.4\textwidth}
    \begin{lstlisting}[style=cppstyle]
signed int a;
a = 10;
signed int *rref;
signed int *return_value;
FUNCTION_CALL:
  return_value = move(&a)
rref = return_value;
assert(*rref == 10);
*rref = 5;
assert(*rref == 5);
    \end{lstlisting}
    \caption{GOTO program of Rvalue reference}
    \label{fig:rvalue_goto}
    \end{subfigure}
\caption{ESBMC verified an example with Rvalue reference}
\label{fig:rvalue_example}
\end{figure*}

\subsubsection{Move Member Functions}

Move constructors and move assignment operators are the main applications of \textit{rvalue} references; they are typically used in classes that manage resources, aiming to optimize resource movement and enhance efficiency. In Clang, when a C++ class or struct does not explicitly define the move semantics member functions, the compiler automatically generates them. The new C++ front-end of ESBMC v7.6 parses these default member functions from the Clang AST, while in ESBMC v2.1, undefined default member functions were not feasible. In line 9 of Figure~\ref{fig:move_code}, we use the default \textit{move} constructor to initialize the struct. Clang's AST provides it since we have not explicitly defined a constructor in the struct.

\begin{figure*}[htbp]
  \centering
    \begin{subfigure}[c]{.42\textwidth}
    \begin{lstlisting}[style=cppstyle]
#include <cassert>
#include <utility>
struct MyStruct {
  int value;
};

int main() {
  MyStruct a{10};
  MyStruct b(std::move(a));
  assert(b.value == 10);
}
    \end{lstlisting}
    \caption{Example of move member functions}
    \label{fig:move_code}
    \end{subfigure}
    \hspace{0.5cm}
    \begin{subfigure}[c]{0.49\textwidth}
    \begin{lstlisting}[style=cppstyle]
MyStruct a;
a={ .value=10 };
MyStruct b;
struct MyStruct * return_value;
FUNCTION_CALL: 
  return_value = move(&a)
FUNCTION_CALL:
  MyStruct(&b, return_value)
assert(b.value == 10);
    \end{lstlisting}
    \caption{GOTO program of move member functions.}
    \label{fig:move_goto}
    \end{subfigure}
\caption{ESBMC verified an example with move member functions.}
\label{fig:move_example}
\end{figure*}

\begin{algorithm}[htbp]
\footnotesize
\algsetup{linenosize=\footnotesize}
\caption{\footnotesize Rvalue References}
\label{pf:rvalue}
\begin{algorithmic}[1]
\STATE Parse rvalue reference variable from Clang AST

\STATE Model the rvalue reference as a \textit{pointer}

\IF{Involves calculations with rvalues or assignments to rvalues}
    \STATE Make special adjustment: dereference the \textit{pointer}
\ENDIF

\end{algorithmic}
\end{algorithm}

\subsection{Exception Handling}

Exception handling is a method that C++ uses to manage runtime errors~\cite{10.5555/1407367}; it helps programs handle errors safely and prevent the program from crashing. This approach involves three main components: (1) the throw statement, which is used to raise an exception, (2) the try block, which contains the code that might throw an exception and directs to the first matching catch statement, and (3) the catch statement, which handles the exceptions raised by the throw statement. In our latest Clang-based C++ front-end, we have redesigned the exception handling mechanism, adopting a method similar to that used in ESBMC v2.1. In the new implementation, the front-end parses these components from the Clang AST, which enables better handling complex constructs, such as nested exceptions.

\begin{figure*}[htbp]
  \centering
    \begin{subfigure}[c]{.44\textwidth}
    \begin{lstlisting}[style=cppstyle]
#include <cassert>
struct Base {};
struct Derived : Base{};

int main() {
  try {
    throw Derived();
  }
  catch(Base) {}
  catch(Derived) {assert(0);}
  return 0;
}
    \end{lstlisting}
    \caption{Example of exception handling}
    \label{fig:catch_code}
    \end{subfigure}
    \begin{subfigure}[c]{0.54\textwidth}
    \begin{lstlisting}[style=cppstyle]
   CATCH tag-Base->1, tag-Derived->2
   Derived tmp;
   THROW tag-Derived, tag-Base: tmp
   CATCH 
   GOTO 3
1: Base
   GOTO 3
2: Derived
   ASSERT false
3: RETURN: 0
    \end{lstlisting}
    \caption{GOTO program of exception handling}
    \label{fig:catch_goto}
    \end{subfigure}
\caption{ESBMC verified an example with exception handling}
\label{fig:catch_example}
\end{figure*}

The GOTO program in Figure~\ref{fig:catch_goto} illustrates how exception handling works. The first \textit{catch} instruction marks the start of the \textit{try} block. This instruction holds the tag assigned to each catch statement and the target location of their respective \textit{catch} blocks. If an exception is thrown, ESBMC follows defined rules to jump to the appropriate \textit{catch} statement, including potentially jumping to an invalid catch that triggers a verification error, indicating that the exception cannot be caught. If a suitable exception handler is found, the \textit{thrown} value is assigned to the \textit{catch} variable if one exists; otherwise, an error will be reported if no valid handler is present. The matching rules for exception handling are listed below:

\begin{enumerate}
  \item \textit{Basic Type:} Exceptions are caught if their type matches the \texttt{catch} type, ignoring qualifiers such as \texttt{const}, \texttt{volatile}, and \texttt{restrict}.

  \item \textit{Array and Pointer:} A pointer type in the \texttt{catch} block can catch exceptions of the corresponding array type.

  \item \textit{Function Pointer:} A \texttt{catch} block for a pointer to a function can catch exceptions of functions with the same return type.

  \item \textit{Base Class:} Exceptions derived unambiguously from the \texttt{catch} block's type are caught.

  \item \textit{Convertible Type:} Exceptions are caught if they can be converted to the type specified in the \texttt{catch} block through standard conversions or qualification adjustments.

  \item \textit{Void Pointer:} A \texttt{void*} in the \texttt{catch} block can catch any pointer type exception.

  \item \textit{Ellipsis:} Any exception can be caught using an ellipsis (\texttt{...}) in the \texttt{catch} block.
  
  \item \textit{Re-throw:} If no new exception is thrown, the last thrown exception should be re-thrown.
\end{enumerate}

We have extended the symbolic engine to improve exception handling in ESBMC v7.6. As illustrated in Figure~\ref{fig:catch_code}, both exception handlers can catch the thrown exception. In ESBMC v7.6, the exception will be caught by \textit{Base} in line 6, stopping the execution of subsequent exception handlers. Therefore, this sample code will not trigger the assertion, and the verification result will be successful. 

As part of the exception handling mechanism, exception specifications clarify a function's exception behavior by defining which exceptions a function can throw. In ESBMC v2.1, we implemented Dynamic Exception Specification, which uses the \textit{throw} keyword to declare a list of exception types that a function can throw, and the first line of Figure~\ref{fig:excpt} illustrates the two types of exceptions that a function is allowed to throw: \textit{int} and \textit{double}. However, with updates to the C++ standard, the Dynamic Exception Specification was deprecated due to its limitations on flexibility. Consequently, we have supported Non-Dynamic Exception Specification in the new front-end. As shown in the third line of Figure~\ref{fig:excpt}, the \textit{noexcept} keyword provides a modern way to declare a function's exception behavior. We used the THROW DECL instruction at the beginning of the function to check if any thrown exceptions match the exception specification. If the thrown exception violates the exception specification, it will result in an assertion property violation.

\begin{figure}[htbp]
  \centering
    \begin{subfigure}{.5\textwidth}
    \begin{lstlisting}[style=cppstyle]
void func() throw(int, double);

void func() noexcept;
    \end{lstlisting}
    \end{subfigure}
\caption{Example of exception specification}
\label{fig:excpt}
\end{figure}

From Process Flow~\ref{pf:exception}, it is clear that exception handling involves parsing the throw statement, try block, and catch block from the Clang AST. If the function includes an exception specification, the thrown exception is verified to conform to the specified constraints. When a throw statement is executed within a try block, a matching catch block is checked. If such a block can handle the exception, the execution jumps to it. Otherwise, an assertion violation is reported for the unhandled exception. Similarly, if a throw statement is executed outside any try block, an assertion violation is raised due to the absence of a handler. Furthermore, if the throw occurs during a function call and is not caught within that function, it will propagate to the calling function.

\begin{algorithm}[htbp]
\footnotesize
\algsetup{linenosize=\footnotesize}
\caption{\footnotesize Exception Handling}
\label{pf:exception}
\begin{algorithmic}[1]
\STATE Parse \textit{throw} statement, \textit{try} block and \textit{catch} block from Clang AST

\IF{Function has exception specification}
    \STATE Check that the thrown exception conforms to the specification
\ENDIF 

\IF{\textit{Throw} statement is executed inside a \textit{try} block}   
    \IF{A \textit{catch} block exists and can catch that exception}
        \STATE Jump to the corresponding \textit{catch} block
    \ELSE
        \STATE Assertion property violation: Failure to catch an exception
    \ENDIF
\ELSIF{\textit{Throw} statement is executed outside the \textit{try} block}
    \STATE Assertion property violation: Failure to catch an exception
\ENDIF

\end{algorithmic}
\end{algorithm}

\subsection{C++ Operational Model}

ESBMC employs an abstract representation of the STL known as the C++ OMs, which are manually created and maintained. These models define function contracts, including pre- and post-conditions, for the STL functions and method calls they encompass, while also having side effects, such as exception propagation. Verifying all these simplifies verification. These OMs were developed based on the old front-end, which utilizes a CPROVER-based parse tree for its type checker. Therefore, the static checking capabilities of these OMs rely on maintenance. As the C++ standard updates, the code within these OMs has gradually become outdated.

With our new Clang-based front-end, static checking has become more compliant with C++ standards. This is due to Clang's following of language standards, advanced type deduction and checking mechanisms. Consequently, ESBMC v7.6 can now detect and report potential program issues during parsing. To adapt the OMs to the new front-end, we identified and addressed parsing errors in the OMs that required fixes, as shown in Table~\ref{table:cpp_libraries}. Consequently, we updated the outdated code syntax, standardized variable names, and improved readability.

\begin{table}[htbp]
\centering
\footnotesize
\begin{tabular}{c p{9cm}}
\hline
\textbf{Category} & \textbf{Operational models} \\
\hline
Containers & \texttt{vector}, \texttt{queue}, \texttt{deque}, \texttt{set}, \texttt{map}, \texttt{iterator}, \texttt{algorithm}, \texttt{stack}, \texttt{bitset} \\
Streams Input/Output & \texttt{istream}, \texttt{ios}, \texttt{ostream}, \texttt{sstream},  \texttt{fstream}, \texttt{streambuf} \\
Strings & \texttt{string}, \texttt{string\_view} \\
Numeric & \texttt{numeric}, \texttt{valarry} \\
Language Support & \texttt{typeinfo}, \texttt{exception} \\
General & \texttt{memory}, \texttt{stdexcept} \\
Localization & \texttt{locale} \\
\hline
\end{tabular}
\caption{Overview of the fixed C++ operational models}
\label{table:cpp_libraries}
\end{table}

\section{Experimental Evaluation}
\label{experimental-evaluation}

We used benchmarks from Monteiro et al.~\cite{monteiro2022model} to evaluate ESBMC v7.6, which were previously used to assess ESBMC v2.1 in the same study.

We used benchmarks to verify the core C++ language features. There are $532$ test cases (TCs) in total over $6$ benchmarks collections. The set of benchmarks \textit{cpp} contains example programs from the book \textit{C++ How to Program}~\cite{10.5555/1407367}.
The inheritance and polymorphism benchmarks are extracted from~\cite{monteiro2022model}. There are three benchmark collections for template specialization -{}- \textit{cbmc-template} comes from the CBMC regressions~\cite{cbmctestsuite}; \textit{gcc-template-tests} were extracted from the GCC template test suite~\cite{gcctestsuite}; \textit{template} is also from benchmarks used in~\cite{monteiro2022model}. The \textit{cpp} set contains programs with mixed use of various C++ language features combined with inheritance, polymorphism, templates, and dynamic memory. Finally, we evaluated the 1001 TCs that depend on the OMs in each benchmark, and these test cases contain the most frequently used STL libraries.

\subsection{Objectives and Setup}
\label{objectives} 

Our evaluation framework is based on \textit{BenchExec}~\cite{10.1007/s10009-017-0469-y}.
For each TC in the test suite, we check whether the verification verdict reported by each tool matches the expected outcome. A TC passes when the tool reports a verdict of ``VERIFICATION SUCCESSFUL'' on a program without any violation of properties or reports ``VERIFICATION FAILED'' on an unsafe program that violates a property. Such properties include arithmetic overflows, array out-of-bounds accesses, memory issues, or assertion failures. Our evaluation aims to answer the following experimental questions: 
\begin{tcolorbox}
\begin{enumerate}
\item[\textbf{EQ1}] (\textbf{soundness}): Can ESBMC v7.6 give more correct verification results and a higher pass rate than its previous versions?
\item[\textbf{EQ2}] (\textbf{performance}): How long does ESBMC v7.6 take to verify C++ programs? 
\item[\textbf{EQ3}] (\textbf{completeness}): Does the tool implement the proposed improvements in complex template support outlined as future work by Monteiro et al.~\cite{monteiro2022model}?
\end{enumerate}
\end{tcolorbox}

The experiment was set up in Ubuntu 22.04 running on an $8$-core Intel CPU and $16$GB RAM, with a time limit of 900 seconds and a memory limit of 6GB. The dataset, scripts, and logs are publicly available on Zenodo~\cite{zenodoarchive}. The cumulative verification time represents the CPU time elapsed for each tool finishing all benchmarks. 

\subsection{Results}
\label{results} 
Table~\ref{table:Evaluation-results} shows our experimental results. With a higher pass rate than ESBMC v2.1 across all benchmarks and outperforming ESBMC v7.3 on $4$ out of $6$ benchmark sets, ESBMC v7.6 successfully achieved more correct results, confirming \textbf{EQ1}. As for ESBMC v2.1, the failed TCs in \textit{cpp} are due to parsing or conversion errors, meaning the previous tool version is unable to properly typecheck the input programs, probably due to the weak parser, as described in Section~\ref{background}. The failed TCs in \textit{inheritance} and \textit{polymorphism} set contain a common feature of dynamically casting a pointer of a child class with a base class containing virtual methods. ESBMC v2.1 could not handle this type of casting, giving conversion errors. 

\begin{tcolorbox}
\textbf{EQ1:} ESBMC v7.6 has a higher pass rate across all benchmarks compared to v2.1 and v7.3, features a more powerful parser to support more C++ features.
\end{tcolorbox}

ESBMC v2.1 has limited support for C++ templates, matching our expectations as reported by Monteiro et al.~\cite{monteiro2022model}. The failed TCs in the \textit{cbmc-template} set are the results of ESBMC v2.1 not able to handle the default template type parameter or explicit template specialization combined with C++ \textit{typedef} specifier.
The low pass rate of ESBMC v2.1 on the \textit{gcc-template-tests} set indicates that this version cannot verify test cases used by an industrial-strength compiler. \textbf{EQ3} is affirmed through the experiment, as none of these problems persist in ESBMC v7.6.

ESBMC v7.3 demonstrates better accuracy than v2.1 in features like inheritance and polymorphism, the failed TCs in the \textit{cpp} and \textit{template} sets are caused by outdated OMs and conversion errors. This indicates a lack of support for certain C++ syntax in the front-end and insufficient consideration of some edge cases during the conversion process. Regarding ESBMC v7.6, it addresses most of the parse and conversion errors in v7.3. However, some TCs fail because Clang generates different ASTs depending on the C++ standard, which the front-end does not yet fully support.

\begin{table}[htbp]
\begin{center}
\footnotesize
\begin{tabular}{c c c c}
\hline 
Benchmarks                 &ESBMC-v2.1  &ESBMC-v7.3   &ESBMC-v7.6 \\ \hline
    cpp                     &{$71\%$}     &{$61\%$}     &{$83\%$}     \\
    inheritance             &{$73\%$}     &{$93\%$}     &{$93\%$}     \\
    polymorphism            &{$80\%$}     &{$84\%$}     &{$85\%$}     \\
    cbmc-template           &{$92\%$}     &{$97\%$}     &{$97\%$}     \\
    gcc-template-tests      &{$39\%$}     &{$71\%$}     &{$78\%$}     \\
    template                &{$53\%$}     &{$65\%$}     &{$81\%$}     \\
    Total verification Time &{$1090s$}     &{$71s$}     &{$545s$}
    \\\hline

\end{tabular}
\end{center}
\caption{Experimental results showing the pass rate for each set of benchmarks and accumulative verification time. This experiment uses ESBMC with Boolector SMT solver.}
\label{table:Evaluation-results}
\end{table}

\begin{figure}[htbp]
    \centering
    \begin{subfigure}[b]{0.43\textwidth}
        \centering
        \includegraphics[width=\textwidth]{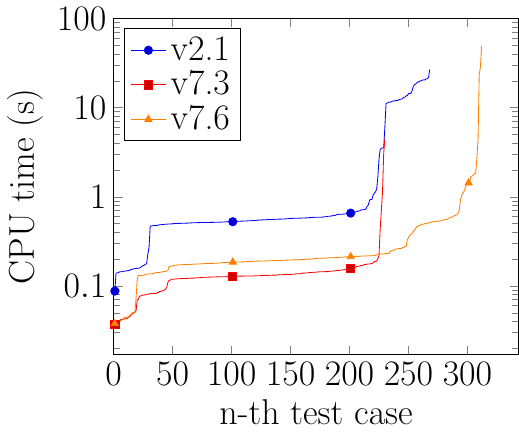}
        \caption{\textit{cpp} set}
        \label{fig:cpp-comp}
    \end{subfigure}
    \hspace{0.05\textwidth}
    \begin{subfigure}[b]{0.43\textwidth}
        \centering
        \includegraphics[width=\textwidth]{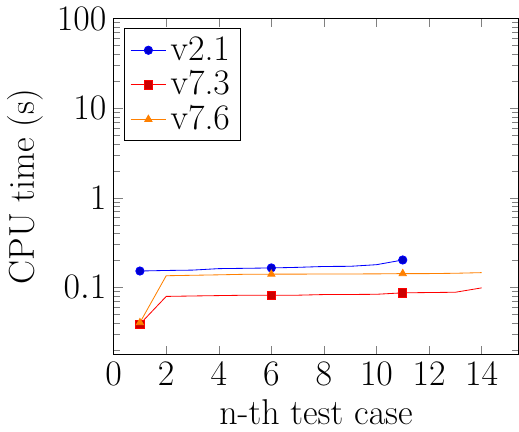}
        \caption{\textit{inheritance} set}
        \label{fig:inh-comp}
    \end{subfigure}
    \begin{subfigure}[b]{0.43\textwidth}
        \centering
        \includegraphics[width=\textwidth]{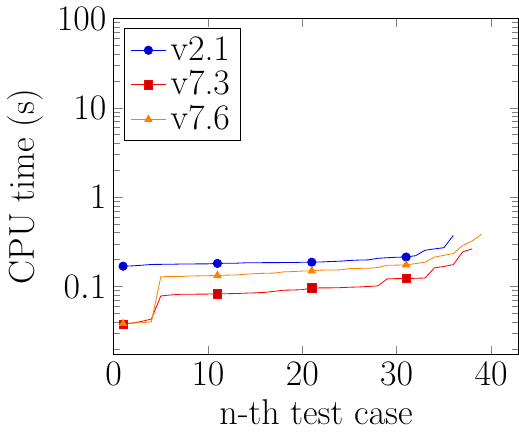}
        \caption{\textit{polymorphism} set}
        \label{fig:poly-comp}
    \end{subfigure}
    \hspace{0.05\textwidth}
    \begin{subfigure}[b]{0.43\textwidth}
        \centering
        \includegraphics[width=\textwidth]{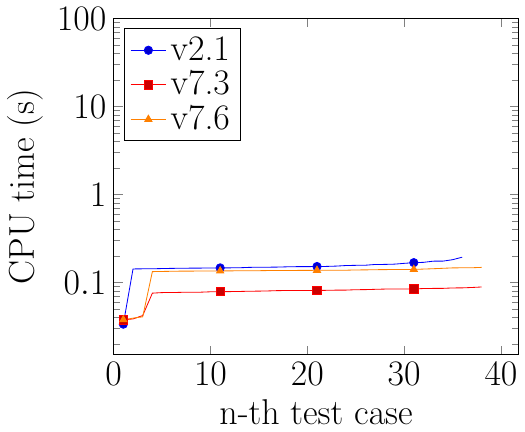}
        \caption{\textit{cbmc-template} set}
        \label{fig:cbmc-comp}
    \end{subfigure}
    \begin{subfigure}[b]{0.43\textwidth}
        \centering
        \includegraphics[width=\textwidth]{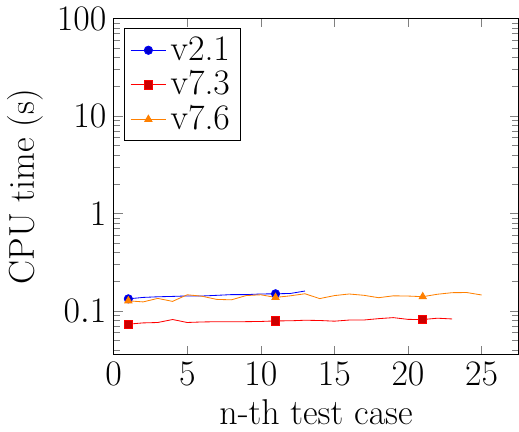}
        \caption{\textit{gcc-template-tests} set}
        \label{fig:gcc-comp}
    \end{subfigure}
    \hspace{0.05\textwidth}
    \begin{subfigure}[b]{0.43\textwidth}
        \centering
        \includegraphics[width=\textwidth]{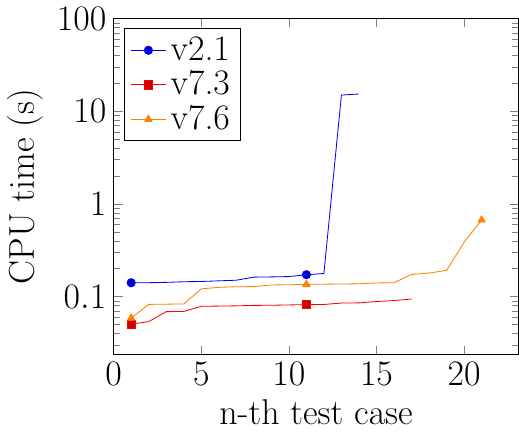}
        \caption{\textit{template} set}
        \label{fig:temp-comp}
    \end{subfigure}
    \caption{CPU time comparison for ESBMC versions on 6 benchmark sets (correct results only)}
    \label{fig:time_comparison}
\end{figure}

Only test cases verified correctly within the time limit are used for performance comparison. There were two timeouts for ESBMC v2.1 and none for v7.3 and v7.6. As illustrated in Figure~\ref{fig:time_comparison}, we conducted experiments on each set of benchmarks. For simple benchmark sets like the \textit{inheritance} and \textit{polymorphism}, the runtime of the three ESBMC versions is comparable. However, for more complex benchmark sets, such as the \textit{cpp} and \textit{template}, the performance of v7.6 and v7.3, which use the Clang-based C++ front-end, shows improvements over v2.1. This is because the Clang-based front-end can efficiently parse complex C++ syntax and generate high-quality AST. The performance decrease from v7.6 to v7.3 results from the front-end's extended support for more advanced C++ features, which increases computational cost.

\begin{tcolorbox}
\textbf{EQ2:} ESBMC v7.6 is more efficient than v2.1, and although slightly slower than v7.3, it delivers better accuracy.
\end{tcolorbox}

\begin{tcolorbox}
\textbf{EQ3:} ESBMC v7.6 handles complex template features more effectively and offers better extensibility.
\end{tcolorbox}

Overall, we have enhanced the template support in ESBMC v7.6, which addresses a key aspect of the future work proposed by Monteiro et al.~\cite{monteiro2022model}. Compared to ESBMC v2.1, ESBMC v7.6 provides faster and more reliable performance. Although v7.6 takes slightly longer than v7.3, it delivers a notable improvement in the accuracy of results.

\begin{figure*}[htbp]
  \centering
  \includegraphics[width=0.45\textwidth]{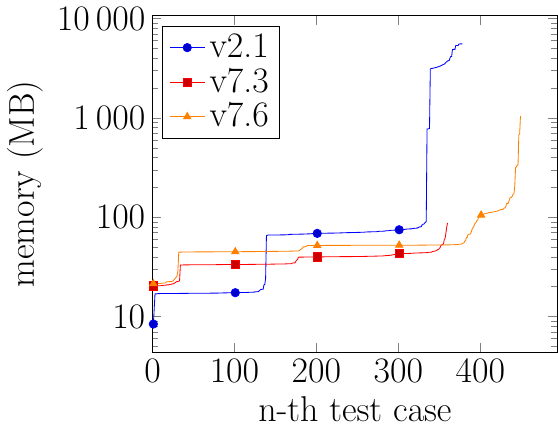}
  \caption{RSS comparison for ESBMC versions on all benchmarks (correct results only)}
  \label{fig:mem-c}
\end{figure*}

In addition to the pass rate and verification time in Table~\ref{table:Evaluation-results}, we assessed each tool's memory usage. Table~\ref{table:memory-usage} shows each benchmark's cumulative maximum RSS (Resident Set Size) using each tool under evaluation. Our metrics collection approach is based on \emph{BenchExec}'s efficient monitoring capabilities provided by the \emph{control} \emph{groups} memory subsystem~\cite{10.1007/s10009-017-0469-y}.
Compared to ESBMC v2.1, ESBMC v7.6 has high pass rates and uses less memory in total. The increased memory usage in v2.1 for the \textit{cpp} and \textit{template} set is because of its inefficiency in verifying test cases involving templates due to its limitations in handling C++ templates and complex features effectively. Many TCs failed due to a CONVERSION ERROR in ESBMC v2.1's front-end and never even reached the solver in the back-end. Figure~\ref{fig:mem-c} shows that v2.1 uses less memory for simple TCs but consumes exponentially more for complex ones. ESBMC with a Clang-based front-end achieves stable and lower memory usage due to its efficient parsing and conversion processes.

\begin{table}[ht]
\begin{center}
\footnotesize
\begin{tabular}{c r r r}
\hline
Benchmarks          &ESBMC-v2.1     &ESBMC-v7.3    &ESBMC-v7.6 \\ \hline
    cpp                 &{$219000$ MB}  &{$13800$ MB}  &{$26900$ MB}  \\
    inheritance         &{$236$ MB}     &{$490$ MB}    &{$653$ MB}    \\
    polymorphism        &{$722$ MB}     &{$1480$ MB}   &{$1960$ MB}   \\ 
    cbmc-template       &{$643$ MB}     &{$1260$ MB}   &{$1680$ MB}   \\
    gcc-template-tests  &{$457$ MB}     &{$1030$ MB}   &{$1390$ MB}   \\
    template            &{$20800$ MB}   &{$691$ MB}    &{$1240$ MB}    \\ 
    Total memory        &{$242000$ MB}  &{$18700$ MB}  &{$33800$ MB}   \\\hline
\end{tabular}
\end{center}
\caption{Experimental results showing each benchmark's cumulative maximum RSS (Resident Set Size). This experiment uses ESBMC with the Boolector SMT solver.}
\label{table:memory-usage}
\end{table}

In ESBMC v2.1, we simulated the behavior of the C++ STL library using OMs and added safety properties. Since then, our C++ front-end has been completely rewritten based on Clang AST, and the back-end has undergone significant development. Comparing v7.6 with v7.3, we have updated these outdated OMs and resolved their issues. We believe it is essential to re-evaluate v7.6 over the C++ STL library benchmarks~\cite{monteiro2022model} using these existing OMs.

As shown in Table~\ref{table:tc}, ESBMC v2.1 has generally high pass rates across most benchmarks, indicating strong support for OMs with security properties. In v7.3, the refactored front-end and outdated OMs resulted in poor pass rates due to the lack of support for several core language features. By comparison, the pass rates for most benchmarks have significantly improved with v7.6, with many returning to or exceeding the pass rates in v2.1. This indicates that the adaptation of the new front-end to the OMs has largely been resolved. Nevertheless, some benchmarks, such as \textit{Multiset}, \textit{Set}, and \textit{Deque}, still lag behind the performance seen in v2.1. Most of the test cases failed due to parsing errors caused by initialization errors in the container OM. This indicates a need for further improvement in our OMs. Additionally, some errors arose from unsupported Clang AST nodes, and extending the front-end to support these AST nodes remains an ongoing development effort.

\begin{table}[htbp]
\begin{center}
\footnotesize
\begin{tabular}{c c c c}
\hline
Benchmarks      & ESBMC-v2.1      & ESBMC-v7.3 & ESBMC-v7.6 \\
\hline
string          & \textbf{99\%}   & 0\%        & 88\%       \\
stream          & \textbf{89\%}   & 33\%       & 88\%       \\
algorithm       & 42\%            & 0\%        & \textbf{80\%}      \\
deque           & \textbf{95\%}   & 0\%        & 88\%       \\
list            & 53\%            & 0\%        & \textbf{65\%}       \\
map             & \textbf{83\%}   & 0\%        & 81\%       \\
multimap        & 89\%            & 0\%        & \textbf{91\%}       \\
multiset        & \textbf{74\%}   & 0\%        & 18\%       \\
priority-queue  & \textbf{100\%}  & 0\%        & 87\%       \\
set             & \textbf{83\%}   & 0\%        & 60\%       \\
stack           & 86\%            & 0\%        & \textbf{86\%}       \\
vector          & 22\%            & 0\%        & \textbf{89\%}      \\
try-catch       & \textbf{88\%}   & 0\%        & 80\%       \\
\hline
\end{tabular}
\end{center}
\caption{Pass rates of OM-dependent benchmarks for C++ STL libraries.}
\label{table:tc}
\end{table}

\subsection{Performance Using Different SMT Solvers}
\label{performance-smt-solvers}

ESBMC v7.6 supports multiple SMT solvers in the back-end, such as Z3~\cite{moura2008z3}, Bitwuzla~\cite{DBLP:conf/cav/NiemetzP23}, Boolector~\cite{brummayer2009boolector}, MathSAT~\cite{bruttomesso2008mathsat}, CVC4~\cite{barrett2011cvc4}, CVC5~\cite{barbosa2022cvc5} and Yices~\cite{dutertre2014yices}. We also evaluated ESBMC v7.6 with various solvers over the same set of benchmarks. Table~\ref{table:results-solvers-time-memory} shows the total verification time and memory consumption for ESBMC v7.6 using different solvers. 

\begin{table}[htbp]
\begin{center}
\footnotesize
\begin{tabular}{c c c}
\hline 
Solvers                 &Time           &Memory\\ \hline
Boolector               &{$545s$}       &{$33800$ MB} \\
CVC4                    &{$2230s$}      &{$44100$ MB} \\
CVC5                    &{$1420s$}      &{$42300$ MB} \\
MathSAT                 &{$2930s$}      &{$50500$ MB} \\
Yices                   &{$1270s$}      &{$40800$ MB}  \\
Z3                      &{$1430s$}      &{$34300$ MB}  \\
Bitwuzla                &{$549s$}       &{$38700$ MB}  \\\hline
\end{tabular}
\end{center}
\caption{Experimental results showing the total verification time and memory consumption for ESBMC v7.6 using different solvers.}
\label{table:results-solvers-time-memory}
\end{table}

Overall, ESBMC v7.6 with Boolector is the fastest configuration that consumes the minimum amount of memory to verify all benchmarks, while Bitwuzla performs similarly but consumes more memory. Among the other solvers, the memory consumption of ESBMC v7.6 with Z3 is close to the Boolector configuration.

\subsection{Threats to Validity}
\label{threats-to-validity}

While developing the new C++ front-end, we encountered challenges in determining the correct order of constructor and destructor calls for the most derived class when analyzing complex inheritance hierarchies in Clang AST, such as the diamond inheritance pattern. We documented it under an umbrella issue currently in our backlog~\cite{esbmccppsupport} on ESBMC GitHub repository~\cite{esbmcissue940}. ESBMC v2.1 mimics the semantics of the APIs of C++ STL libraries using a set of OMs. The C++ front-end of ESBMC has been completely rewritten, and the back-end has also undergone significant development and evolution since v2.1 was published in~\cite{monteiro2022model}. Additionally, the number of these OMs is large, and for libraries without added safety properties, using the C++ standard library directly is the best solution. However, it is uncertain whether ESBMC's C++ front-end can fully support the standard library.

\section{Conclusions and Future Work}
\label{conclusion-and-future-work}

We present a new Clang-based front-end that converts in-memory Clang AST to ESBMC's IR. In our evaluation of ESBMC v7.6, we compared it to ESBMC v2.1 and v7.3, specifically focusing on benchmarks to cover core C++ language features. The results demonstrate significant progress with ESBMC v7.6, as it successfully handles real-world C++ programs, including those from the GCC test suite. Notably, it significantly reduces the number of conversion and parse errors compared to the previous version, showcasing improved performance over the benchmarks for core language features. 

While ESBMC effectively mimics the semantics of APIs of the STL libraries using the OMs from ESBMC v2.1, we recognize the need for continuous improvement. As we endeavor to verify modern C++ programs, these OMs require regular review and updates to align with the C++ standard used in the input program. Accurate OMs are essential, as any approximation may lead to incorrect encoding and invalidate the verification results. With ESBMC v7.6, we improved the front-end, updated the OMs, and added support for more core C++ language features. Overall, while our Clang AST-based C++ front-end has not fully restored or improved performance across all benchmarks, the experimental results show substantial improvements compared to previous versions. This highlights the potential of the new front-end.

Furthermore, as part of the remaining future work from Monteiro et al.~\cite{monteiro2022model}, our OMs have yet to support certain C++11 features, including new sequential and unordered associative containers, as well as multithreaded libraries, which remain areas for future development. Our previous success verifying a commercial C++ telecommunication application using ESBMC v2.1 has inspired further goals~\cite{SousaCF15,monteiro2022model}. With ESBMC v7.6 and beyond, we plan to verify the C++ interpreter in OpenJDK as part of the Soteria project~\cite{soteriaorg}.


\section{Acknowledgements}
\label{acknowledge}

The ESBMC development is currently funded by ARM, Intel, EPSRC grants EP/T026995/1, EP/V000497/1, EU H2020 ELEGANT 957286, and Soteria project awarded by the UK Research and Innovation for the Digital Security by Design (DSbD) Programme.



\bibliographystyle{elsarticle-num} 
\bibliography{refs}





\end{document}